\documentstyle[aps,epsf,12pt,tighten,titlepage]{revtex}

\mediumtext
\def\dilog{\mathop{\rm Li_2}\nolimits}
\newcommand{\be}[1]{\beta^#1}
\newcommand{\rz}{\rho_Z}
\newcommand{\rw}{\rho_W}
\newcommand{\rh}{\rho_H}

\begin{document}

\begin{titlepage}

\hfill \parbox{5cm}{\raggedleft TTP95--05\\ hep-ph/9502375 \\February 1995}

\begin{center}{\Large \bf QCD Corrections to Electroweak Annihilation Decays of
        Superheavy Quarkonia\footnote{The complete paper, including
	figures, is also available via anonymous ftp at
	\mbox{ttpux2.physik.uni-karlsruhe.de} (129.13.102.139) as
	/ttp95-05/ttp95-05.ps, or via www at
	\mbox{http://ttpux2.physik.uni-karlsruhe.de/preprints.html}} }
\\[3\baselineskip]

J.\,H. K\"uhn and M. Peter \\

{\em Institut f\"ur Theoretische Teilchenphysik\\
        Universit\"at Karlsruhe \\
        D--76128 Karlsruhe, Germany}
\end{center}

\vspace{2\baselineskip}

\begin{abstract}
QCD corrections to all the allowed decays of superheavy groundstate quarkonia
into electroweak gauge and Higgs bosons are presented. For quick estimates,
approximations that reproduce the exact results within less than at worst
two percent are also given.
\end{abstract}

\end{titlepage}

\vspace*{\textheight}
\thispagestyle{empty}
\newpage
\setcounter{page}{1}

\section{Introduction}

The search for new quark flavors remains an important and interesting
task. The existence of a fourth generation cannot be excluded on the basis
of present measurements. The $\rho$-parameter only restricts the possible
mass-splitting and the limit on the `number of neutrinos' refers to light ones
only. In fact, some GUTs even require quarks that do not fit into the standard
model scheme.

Because charm as well as bottom have been discovered through hadronic
production of their quarkonium bound states $J/\psi$ and $\Upsilon$
respectively, some authors \cite{barger,mirkes} examined the prospects for
discovering new flavors at future hadron colliders through a similar
mechanism. Given favorable circumstances, gluon fusion would be the dominant
source for quarkonium production and therefore especially the pseudoscalar
ground state $\eta$ could be produced with sufficient rate. Also the $^3S_1$
state $\psi$ might be accessible.

For the lighter member of a fourth generation doublet the single quark decay,
i.e.~the decay of one of the constituents in the quarkonium, is likely to be
suppressed due to small intergeneration mixing angles. Thus one may in a first
step ignore this channel and only consider the annihilation decays. The latter
include some new and distinctive modes which may even become dominant. The
most important one could be $\eta\to ZH$, which might even offer a way to
discover the Higgs boson \cite{barger,barger2}.

Predictions for the decays $\eta,\psi\to\gamma\gamma,\gamma Z,ZZ,WW,ZH$ have
been derived in Born approximation in references \cite{acta,top,barger}. QCD
corrections, however, are only partly known. The aim of this work is to fill
this gap.

The paper is organized as follows: The calculational method employed can be
found in \cite{ego} and will be discussed only briefly in the following
section II, together with some general considerations.  In section III the
results of our calculations will be presented. Compact approximations will be
given in section IV, a brief summary in section V.

\section{General considerations}
Not yet discovered quarks must be heavy and a nonrelativistic treatment of
their bound states is adequate. As is well known, in this case the decay width
of an S-wave bound state factorizes into a nonperturbative part -- the
wavefunction at the origin -- and a perturbative part which is proportional to
the free quark scattering cross section:
\begin{equation}
  d\Gamma(S\to p_1+p_2)=\frac{1}{4m^2}\frac{|R_S(0)|^2}{4\pi}|{\cal M}
              (\vec v=0)|^2d_{Lips}
\end{equation}
where $\vec v$ denotes the relative velocity of the $q\bar q$ system and $\cal
M$ the free scattering amplitude. The factorization in this form applies for
S-states only (for P-states see \cite{ego}) and to the order considered in this
paper. Relativistic corrections first enter at ${\cal O}(\alpha_S^2)$, not
considered in this work.

There are several ways to calculate the rate: one is to simply compute the spin
averaged cross section (with a modified statistical factor $1/4\to1/(2S+1)$
where $S$ refers to the spin of the bound state), if only the desired spin
configurations can contribute to the sum. This may require care concerning
possible couplings. Consider for example the $\gamma H$-mode: among the S waves
only the spin 1 state with $J^{PC}=1^{--}$ can decay that way, and the
approach is straightforward. However, if we switch to the $ZH$-mode, both
states with $J^{PC}=0^{-+}$ and $J^{PC}=1^{--}$ are present in the sum, albeit
with unknown relative weight. In this case one may identify the parity
violating part $\propto a\cdot v$ in the amplitude with the $\eta$ decay, the
part $\propto (v^2+a^2)$ with $\psi$. A second way is to project the
appropriate amplitudes using the method derived in \cite{ego}. Both methods
were used to obtain and check the results given below, with the exception of
the decays into two $W$s. There only the second one was applied because the
separation of the various couplings is inconvenient.

The calculation of the wavefunction $R_S$ requires the knowledge of the QCD
potential. To get rid of the dependence on the potential model all widths are
normalized to
\begin{equation}
   \Gamma(\eta\to\gamma\gamma)=12|R_S(0)|^2\frac{\alpha^2Q^4}{M^2}\label{norm}
\end{equation}
and only the ratios
\begin{equation}
   R^X_{ab} \equiv \frac{\Gamma(X\to ab)}{\Gamma(\eta\to\gamma\gamma)}
\end{equation}
are presented.

\begin{figure}
  \begin{center}
  \epsfxsize 11cm \leavevmode\epsfbox{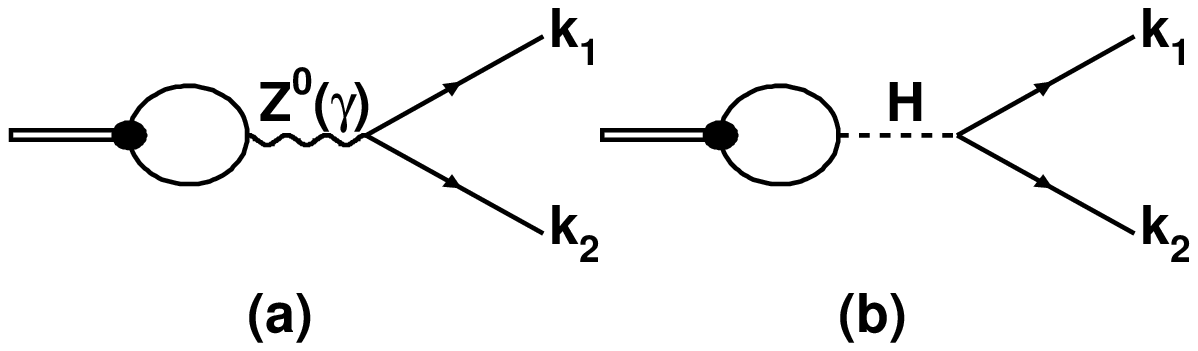}\end{center}\begin{center}
  \epsfxsize 11cm \leavevmode\epsfbox{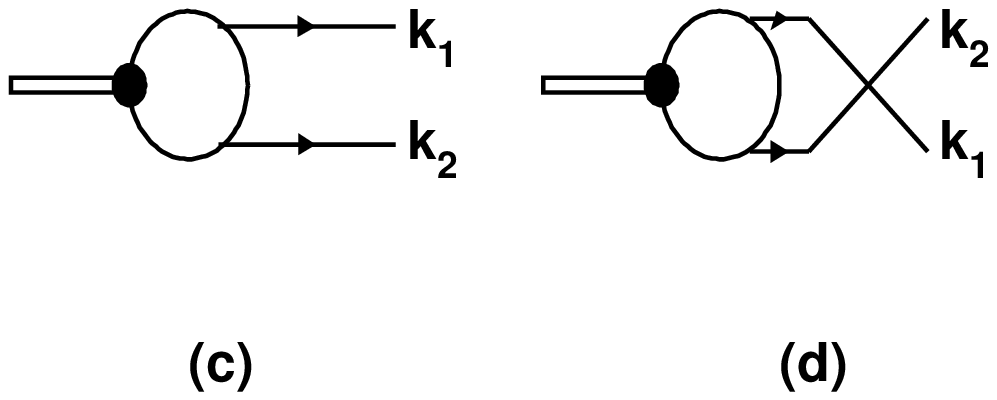}
  \end{center}
  \caption{Tree level generic Feynman diagrams}
  \label{borns}
\end{figure}

The zeroth order generic Feynman diagrams responsible for the annihilation
decays are shown in Fig.~\ref{borns}. The decay through the virtual photon or
$Z$ (Fig.~\ref{borns}(a)) contributes only to the channel $W^+W^-$, the decay
through the virtual Higgs (Fig.~\ref{borns}(b)) is shown only for completeness.
It contributes neither to $\eta$ nor to $\psi$ decays because of its quantum
numbers (this only applies to a standard model Higgs, of course).

\begin{figure}
   \begin{center}
   \epsfxsize 11cm \leavevmode\epsfbox{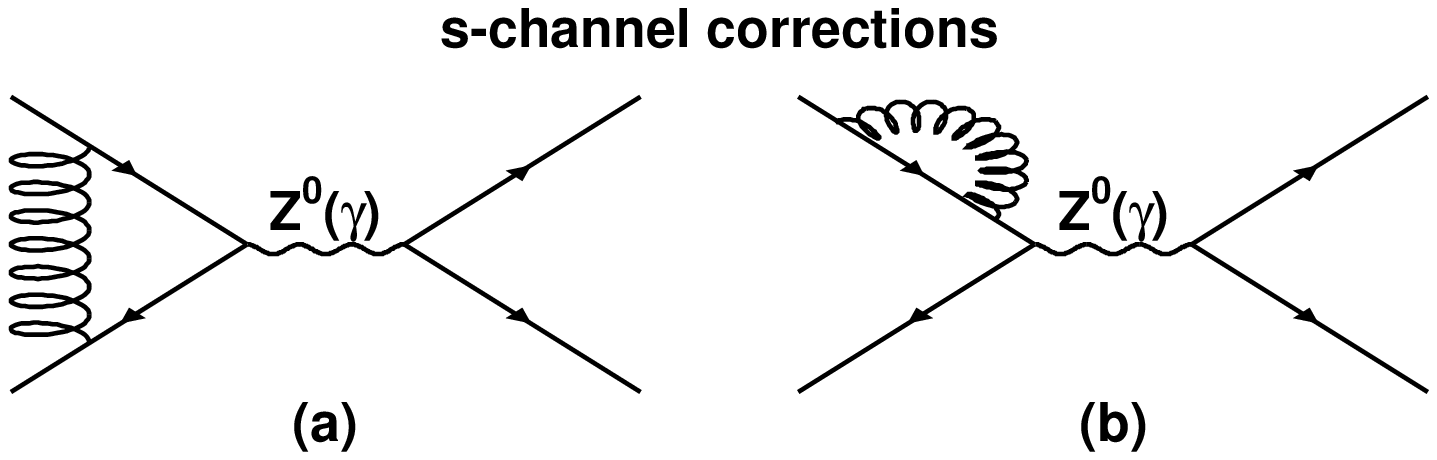}\end{center}\begin{center}
   \epsfxsize 11cm \leavevmode\epsfbox{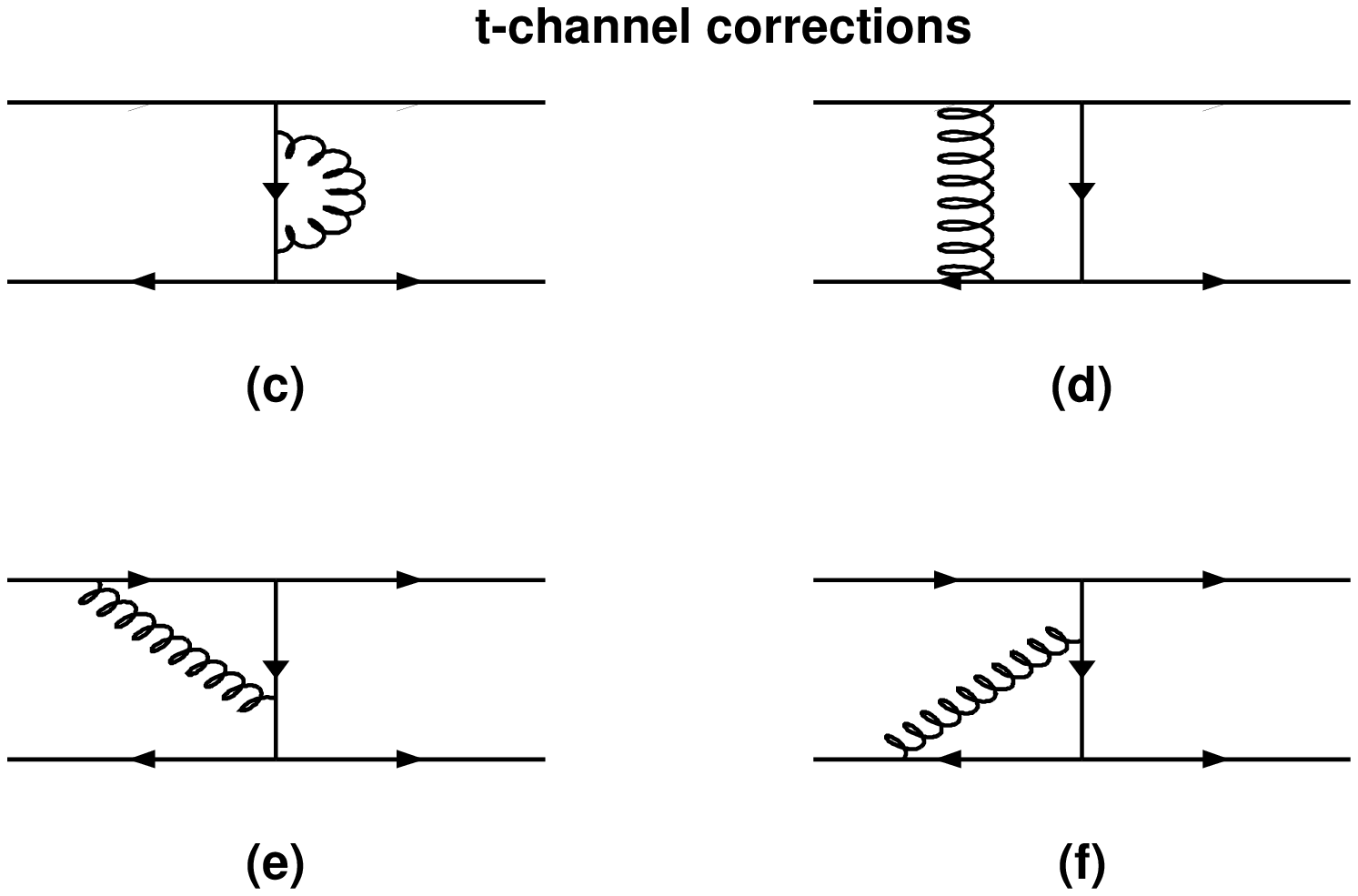}
   \end{center}
   \caption{Diagrams contributing to first order QCD corrections}
   \label{qcds}
\end{figure}

First order QCD corrections receive contributions from the diagrams shown in
Fig.~\ref{qcds}. Their sum is infrared finite. Real gluon emission is
forbidden by color conservation. Comparing this result with the calculation of
QED corrections, which is practically the same in this order, we could also
argue that the coupling of real gluons (photons) to a color (electrically)
neutral state vanishes in the static limit, thus providing an infrared finite
answer.

However, as expected, all corrections exhibit the Coulomb singularity, e.g.
\begin{equation}
  \Gamma(\eta\to\gamma\gamma)=\Gamma^{(Born)}(\eta\to\gamma\gamma)\left(1+
    \frac{\alpha_sC_F}{2\pi}\left(\frac{\pi^2}{|\vec v|}+\frac{\pi^2}{2}-10
    \right)\right)
\end{equation}
which is universal, proportional to $1\over|\vec v|$ ($C_F=4/3$ is a color
factor), and which originates from box- and from s-channel vertex correction
diagrams (Figs.~\ref{qcds}(a) and (d)). This divergence actually represents
part of the Bethe-Salpeter wavefunction and must be dropped since it is
already included in the factor $R_S(0)$. Furthermore, since we are calculating
ratios of decay widths, the singularities (formally) cancel anyway. The
K-factors for the ratios $R$, and their nontrivial parts $\delta k$, which are
defined through

\begin{equation}
  K^X_{ab} \equiv \left(R^X_{ab}\right)_{corrected}:\left(R^X_{ab}\right)_{
     Born} = 1+\frac{\alpha_sC_F}{2\pi}\cdot\delta k^X_{ab}
\end{equation}
will thus be free from Coulomb singularities. The corrected rate can then be
obtained from
\begin{equation}
  \Gamma^X_{ab}=\Big(R^X_{ab}\Big)_{Born}\cdot\Gamma^{Born}(\eta\to\gamma
     \gamma)\cdot\Big(1+\frac{\alpha_SC_F}{2\pi}\Big(\frac{\pi^2}{2}-10+
     \delta k^X_{ab}\Big)\Big)
\end{equation}

At this point a comment on the regularization the Coulomb singularity is
appropriate. Two different procedures are possible: one may either start with
nonvanishing $\beta$ and consider the limit $\beta\to0$ in the end, which
obviously requires significantly more effort during the calculation than really
needed, in particular for the box diagram. Alternatively one may set $\beta=0$
from the outset and employ a nonvanishing gluon mass $\lambda$. This second
procedure has several advantages: the Coulomb singularity and the infrared
divergences are regularized in one step and the special kinematical situation
facilitates the calculation (especially of the box diagram) significantly. To
connect the two approaches the vertex correction can be investigated. This
leads to the substitution rule
\[ \frac{m}{\lambda} \widehat = \frac{\pi}{4|\vec v|}. \]

Before the results of the calculations can be presented, the notation must be
fixed. Vector and axial vector coupling constants are abreviated with the
help of
\[ v=2(I_{3L}+I_{3R})-4Q\sin^2\Theta_w \quad,\quad a=2(I_{3L}-I_{3R})
   \quad,\quad y=2\sin2\Theta_W  \]
where $Q$ denotes the quark charge divided by the proton charge, $I_3$ (the
3-component of) the weak isospin and $\Theta_w$ the weak mixing angle. The
results are applicable to fourth generation quarkonia as well as to more
unconventional quarks (for example an isosinglet in $E_6$ models) as long as
the couplings to the Higgs boson coincide with those of the standard model.

All masses are measured in units of the quarkonium mass $M=2m_q$, whence
$\rho_X=M_X^2/M^2$.

\section{Results}

\subsection{Decays into $ZZ$ or $\gamma\gamma$}
Decays into two Z bosons are possible if the mass $M$ of the quarkonium is
larger than twice the Z mass and proceed via diagrams \ref{borns}(a,c,d).
$\psi\to\gamma\gamma$ is forbidden by C conservation, whereas $\psi\to ZZ$ is
allowed through the parity violating coupling. $\Gamma (\eta\to\gamma\gamma)$
can be obtained from $\Gamma(\eta\to ZZ)$ by replacing $a\to0,~v\to Qy$ and
taking the limit $\rz\to0$, which leads to the result given above
(\ref{norm}). The lowest order predictions for the ratios
\begin{eqnarray}
  R^\eta_{ZZ} & = & \frac{(v^2+a^2)^2}{(Qy)^4}\frac{\be3}{
                    (1-2\rho_Z)^2} \\
  R^\psi_{ZZ} & = & \frac{2}{3}\frac{(av)^2}{(Qy)^4}\frac{\be5}
                    {\rho_Z(1-2\rho_Z)^2}
\end{eqnarray}
where $\beta=\sqrt{1-4M_Z^2/M^2}$ in this case, are well known. To include the
QCD corrections, they have to be multplied by the K-factors, which contain as
their nontrivial parts:
\begin{eqnarray}
  \delta k^\eta_{ZZ} & = & -\frac{1-\be2}{\be2}-\frac{\pi^2}{2}+
        \frac{1+3\be2-5\be4+\be6}{\be4}\ln(1+\be2)-8\frac{2+5\be2-\be4}
        {\be2(1+\be2)}B(\rz)
        \nonumber \\ && -2\frac{1+\be2}{\be2}C(\beta)+\frac{8a^2(1+\be2)}
        {(v^2+a^2)\be2}\Bigg(\ln(1+\be2)+\frac{8}{1+\be2}B(\rz)+C(\beta)
        \Bigg) \\
  \delta k^\psi_{ZZ} & = & 2-\frac{\pi^2}{2}-\frac{2-3\be2+4\be4+\be6}{\be4}
         \ln(1+\be2)-4\frac{6+7\be2+11\be4+2\be6}{\be4(1+\be2)}B(\rz)
         \nonumber \\ && -\frac{6+\be2-9\be4}{2\be4}(1+\be2)C(\beta)
\end{eqnarray}
\begin{eqnarray*}
   B(\rz) & = & \sqrt{\rz(1-\rz)}\arcsin\sqrt{\rz} \\
   C(\beta) &=& -\frac{1}{2\beta}\Bigg\{\frac{\pi^2}{4}+\dilog\Bigg[-\Bigg(
     \frac{1-\beta}{1+\beta}\Bigg)^2\Bigg]-\dilog\Bigg[\Bigg(\frac{1-\beta}
     {1+\beta}\Bigg)^2\Bigg] \\ && \quad +
      2\Re\Bigg[\dilog\Bigg(\frac{1-\beta}{1-i\beta\sqrt{\frac{3+\be2}{1-\be2}
        }}\Bigg)-\dilog\Bigg(\frac{1+\beta}{1+i\beta\sqrt{\frac{3+\be2}{1-\be2}
        }}\Bigg)\Bigg]\Bigg\}
\end{eqnarray*}
$\dilog(x)$ denotes the dilogarithm: $\dilog(x)=-\int_0^1dt\frac{\ln(1-xt)}
{t}$.

\begin{figure}\begin{center}
  \epsfysize 14cm \leavevmode\epsfbox{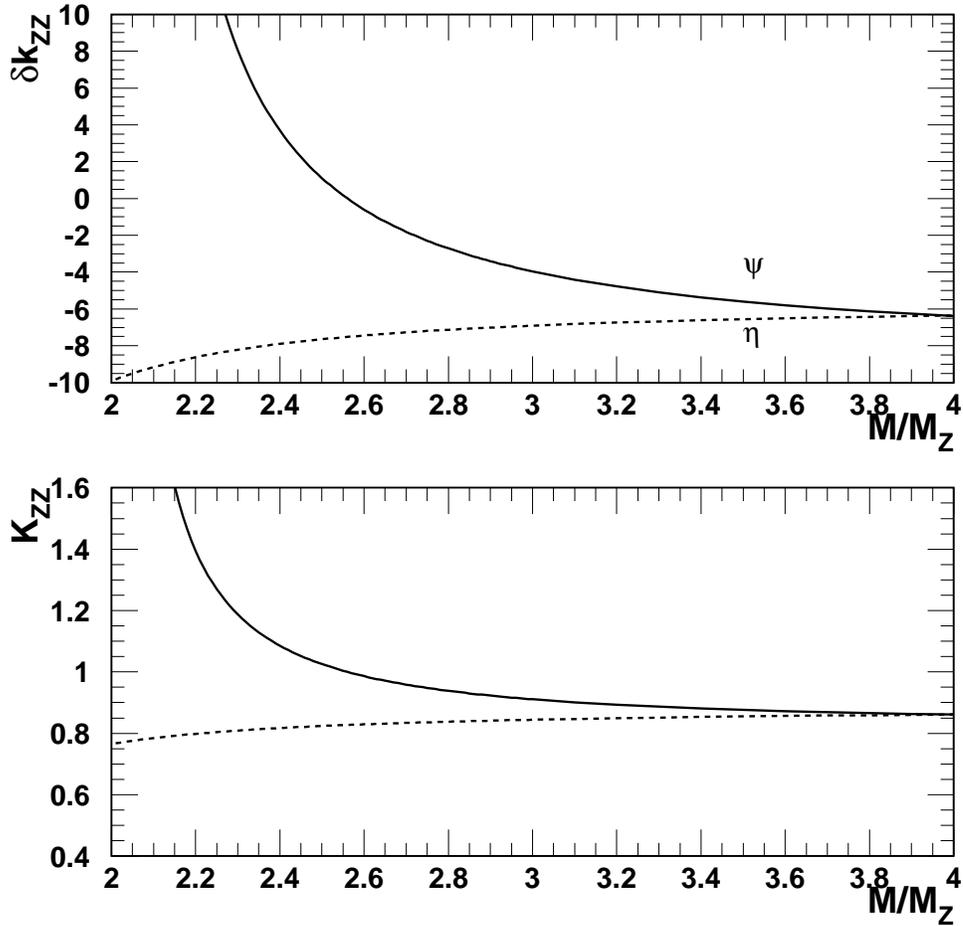}\end{center}
  \caption{$K_{ZZ}$ and $\delta k_{ZZ}$\label{kzz}}
\end{figure}

The limiting case of a large quarkonium mass as well as shorter approximate
formul\ae\ are presented in section IV.

The two K-factors are shown in Fig.~\ref{kzz}, using \cite{mirkes}:
\[ \alpha_S(M^2)=\frac{12\pi}{21\ln(M^2/\Lambda^2)}\quad\mbox{with}\quad
   \Lambda=58MeV\]

The leading apparent singularities of $K^\psi_{ZZ}$ and $K^\eta_{ZZ}$ at
threshold ($M=2M_Z$ or $\beta=0$) proportional to $1/\be4$ and $1/\be2$
respectively cancel. Hence $K^\psi_{ZZ}$ diverges $\propto 1/\be2$,
$K^\eta_{ZZ}$ remains finite:
\begin{eqnarray}
  \delta k^\eta_{ZZ} &\to& \frac{29}{6}-\frac{\pi}{\sqrt{3}}-\frac{\pi^2}{2}
     +\frac{1-\sqrt{3}\pi}{3}\frac{8a^2}{v^2+a^2} \qquad \beta\to0 \\
  \delta k^\psi_{ZZ} &\to& \frac{\sqrt{3}\pi}{\be2}+\frac{16}{3}-\frac{43\pi
      }{5\sqrt{3}}-\frac{\pi^2}{2}\qquad \beta\to0
\end{eqnarray}

The D-wave phase space $\propto\be5$ for $\psi\to ZZ$ which is present in Born
approximation is thus modified to a behavior $\propto\be3$ characteristic for
P-wave phase space.


\subsection{Decays into $Z\gamma$}

The decay to a Z boson plus a photon is possible if the quarkonium mass exceeds
the Z mass. The t- and u-channel quark exchange diagrams (c) and (d)
contribute. The normalized decay rates are given by
\begin{eqnarray}
  R^\eta_{Z\gamma} & = & 2\frac{v^2}{(Qy)^2}\beta \\
  R^\psi_{Z\gamma} & = & \frac{2}{3}\frac{a^2}{(Qy)^2}\frac{\beta(2-\beta)}{
                         (1-\beta)^2}
\end{eqnarray}
(with $\beta=1-M_Z^2/M^2$) and the K-factors by
\begin{eqnarray}
  \delta k^\eta_{Z\gamma} & = & 2\frac{1-\beta}{1-2\beta}+\frac{7\pi^2}{6}
     \frac{1-\beta}{\beta}+\frac{4\beta-3\be2-1}{(1-2\beta)^2}\ln(2\beta)
     \nonumber\\ &&
     -\frac{12}{\beta}B(\rz)-2\frac{3-2\beta}{\beta}\arcsin^2\sqrt{\rz}
     +2\frac{1-\beta}{\beta}\dilog(1-2\beta) \\
  \delta k^\psi_{Z\gamma} & = & -\frac{2\beta}{1-2\beta}+\frac{\pi^2}{6}\frac{
     16-24\beta-5\be2+7\be3}{\be2(2-\beta)}-2\frac{8-39\beta+63\be2-32\be3
     +2\be4}{\beta(2-\beta)(1-2\beta)^2}\ln(2\beta) \nonumber \\ &&
     -4\frac{8-5\beta-\be2}{\be2(2-\beta)}B(\rz)-2\frac{8-13\beta+\be2+2\be3}
     {\be2(2-\beta)}\arcsin^2\sqrt{\rz} \nonumber \\ &&
     +\frac{8-15\beta+2\be2+2\be3}{\be2(2-\beta)}\dilog(1-2\beta)
\end{eqnarray}
with $B(\rz)=\sqrt{\rz(1-\rz)}\arcsin\sqrt{\rz}$ as before.

\begin{figure}\begin{center}
  \epsfysize 14cm \leavevmode\epsfbox{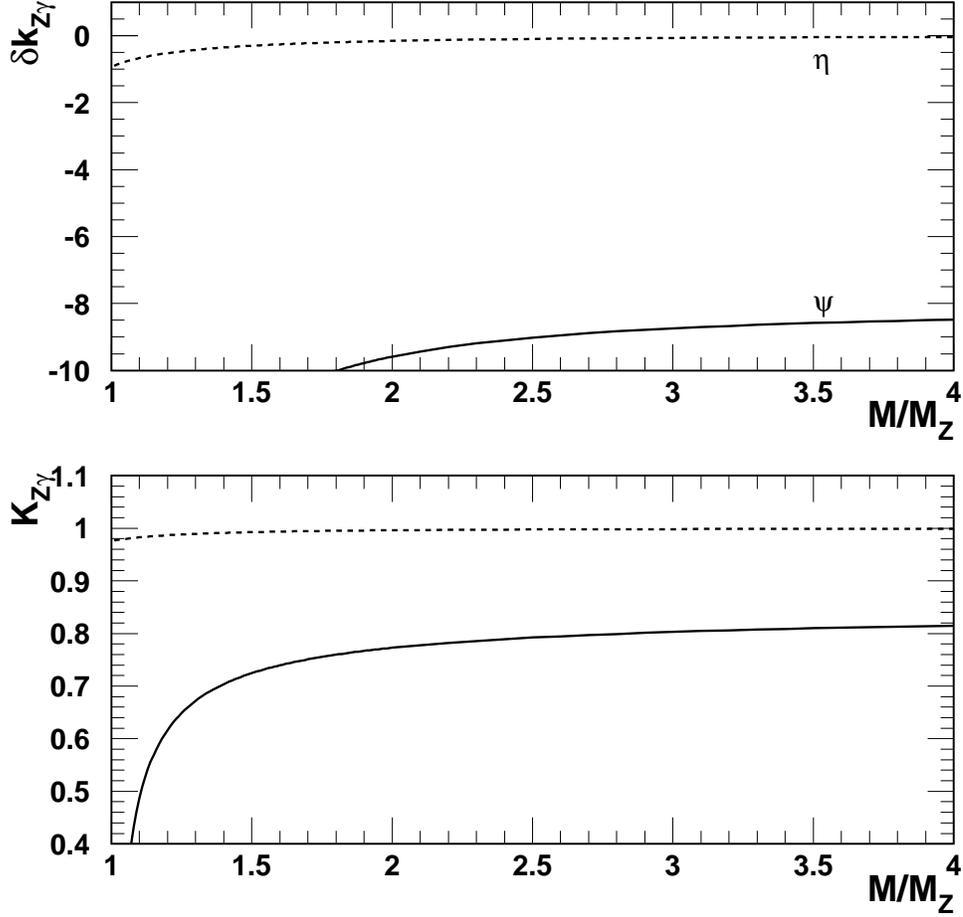}\end{center}
  \caption{$K_{Z\gamma}$ and $\delta k_{Z\gamma}$\label{kzg}}
\end{figure}

These two functions are shown in Fig.~\ref{kzg}.

Close to threshold $\beta$
approaches zero and $\delta k^\psi$ diverges, in contrast to $\delta
k^\eta$. Specifically:
\begin{eqnarray}
   \delta k^\psi_{Z\gamma} \to & -\Bigg(\frac{8\pi}{3}\Big(\frac{1}{\beta}
   \Big)^{1/2}+\frac{\pi^2}{2}-6\Bigg) &,~\beta\to0\\
   \delta k^\eta_{Z\gamma} \to & -\Bigg(\frac{\pi^2}{2}-4\Bigg) &,~\beta\to0.
\end{eqnarray}

This behavior results from the combination of the singularities of the gluon
and the quark propagator. In fact, the virtual quark is close to its mass
shell in this limit. Very close to threshold intermediate bound states should
be taken into account since an alternative way to describe the decay is
through the chain $1^{--}\to\gamma1^{++}(\to Z)$ with mixing between $1^{++}$
and the $Z$. A similar problem also appears in $\psi\to \gamma H$
\cite{bern}. For $\beta=1/2$ the correction is regular.


\subsection{The decays into $ZH$ and $\gamma H$}

As stated in the introduction the decay $\eta\to ZH$ could signal the presence
of the Higgs boson and the new quark simultaneously. In Born approximation
one obtains for the ratios $R$:
\begin{eqnarray}
  R_{ZH}^\eta &=& \frac{a^2}{(Qy)^4}\frac{\be3}{\rz^2} \\
  R_{ZH}^\psi &=& \frac{2v^2}{3(Qy)^4}\frac{\beta}{\rz(1-\rz-\rho_H)^2}
    \nonumber \\ && \times
  \left[\left(1-\rz-\rho_H\frac{1-3\rz}{1-\rz}\right)^2+\frac{\rz}{2}
    \left(1-\rz+\rho_H\frac{2-\rho_H}{1-\rz}\right)^2\right]
\end{eqnarray}
with $\beta=\sqrt{1+\rz^2+\rho_H^2-2\rz-2\rho_H-2\rz\rho_H}$.
In principle, all three diagrams \ref{borns}(a,c,d) could contribute to both
decays. However, the contribution from (c) and (d) vanishes for $\eta\to ZH$
(this is no longer true for the QCD corrections), whereas $R^\psi_{ZH}$ is
the sum of three terms:
\begin{eqnarray}
  R_{ZH,t}^\psi &=& \frac{v^2}{(Qy)^4}\frac{2\beta}{3\rz}\frac{\be2+6\rz}
    {(1-\rz-\rho_H)^2}~\mbox{ t-channel, (c)+(d)} \nonumber \\
  R_{ZH,s}^\psi &=& \frac{v^2}{(Qy)^4}\frac{\beta}{3}\frac{\be2+12\rz}
    {(1-\rz)^2}~\mbox{ s-channel, (a)} \label{zhs}\\
  R_{ZH,st}^\psi &=& -\frac{v^2}{(Qy)^4}\frac{8\beta}{3}\frac{1+\rz-\rho_H}
    {1-\rz+\rho_H}~\mbox{ interference term} \nonumber
\end{eqnarray}

For large quarkonium masses, $R^\eta_{ZH}$ rises proportional to $1/\rz^2$, in
contrast to all other modes listed up to now, which increase proportional to
$1/\rz$ or approach a constant value. The different behavior can be understood
from the Goldstone boson equivalence theorem \cite{gbet} which identifies the
amplitudes for scattering of longitudinal gauge bosons ($Z$ or $W$) at high
energies with those for the appropriate Goldstone bosons. The latter must be
taken as pseudoscalar particles. The decay $\eta\to G_ZH$ ($G_Z$ denotes the
Goldstone boson belonging to $Z$) is allowed, i.e.~the $Z$ emitted can be
longitudinal, with its coupling proportional to $M/M_Z$. This explains one
factor $1/\rz=M^2/M_Z^2$. The second one originates from the coupling of the
Higgs to fermions. In contrast the decay $\psi\to G_ZH$ is forbidden by CP
conservation, and only the transverse part of $Z$ with a coupling proportional
to $g_{weak}$ remains. This leads to a rate $\propto1/\rz$.

The behavior of the other rates can be obtained through similar arguments:
$\eta\to G_ZG_Z,G_ZZ$ or $G_Z\gamma$ are forbidden, hence $R^\eta_{ZZ}$ and
$R^\eta_{Z\gamma}$ approach constant values. The decays $\psi\to G_ZZ$ (but
not $G_ZG_Z$) and $G_Z\gamma$ are allowed, these ratios increase
$\propto1/\rz$.

The full expressions for the K-factors for $R_{ZH}$ can be found in the
appendix. The numerical results are shown in Fig.~\ref{kzh}. The corrections
are large, at least about 15\% for $R^\eta_{ZH}$ and 30\% for $R^\psi_{ZH}$,
but of course they do not drastically change the conclusions that can be
drawn from the Born results, at least if we are not too close to threshold.

\begin{figure}\begin{center}
   \epsfysize 14cm \leavevmode\epsfbox{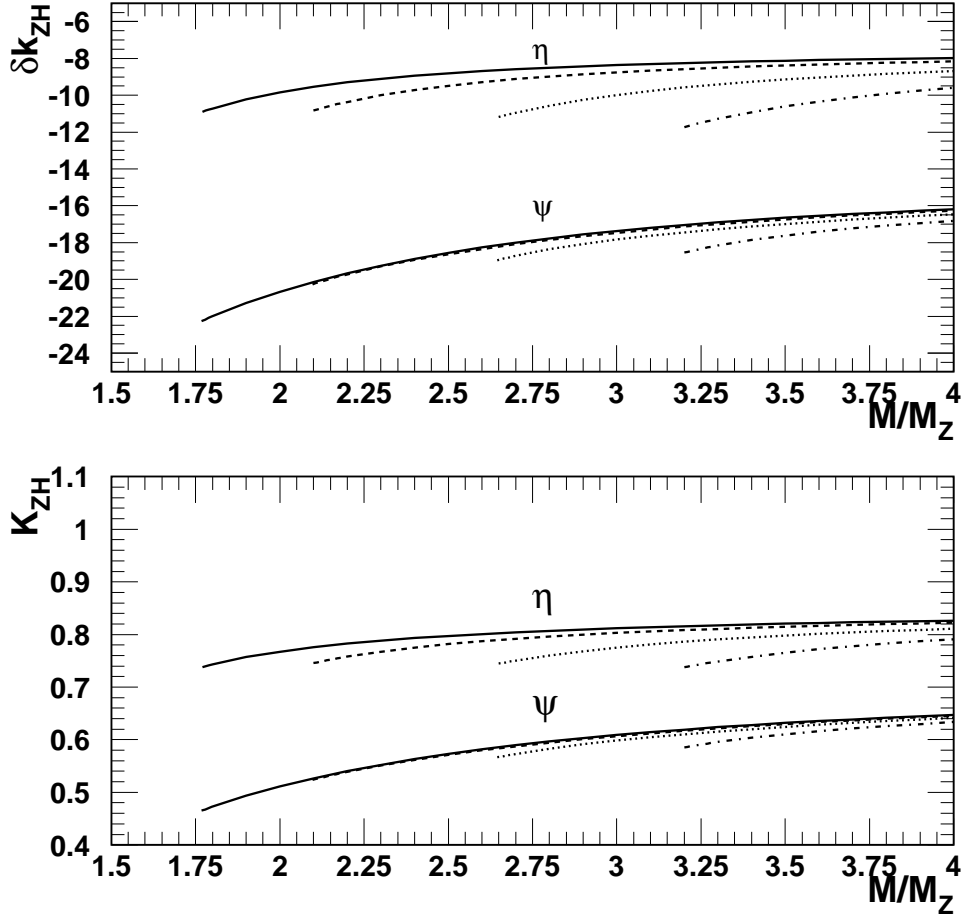}\end{center}
  \caption{$K_{ZH}$ and $\delta k_{ZH}$ for Higgs masses
    $M_H=70,100,150,200GeV$ (solid, dashed, dotted, dash-dotted).\label{kzh}}
\end{figure}

In the limit $\rz\to0$, keeping $\rho_H=M_H^2/M^2$ fixed and also keeping a
factor $1/\rz=(yg_H/e)^2$ in the numerator, the well-known result for the
decay $\psi\to\gamma H$ is recovered \cite{Hg}:
\begin{eqnarray}
   R^\psi_{\gamma H} & = & \frac{1}{(Qy)^2}\frac{2}{3\rz}\Big(1-\rho_H\Big)\\
   \delta k^\psi_{\gamma H} & = & 2\frac{1-3\beta}{1-2\beta}-\frac{\pi^2}{2}
        -8\frac{2-\be2}{\be2(1-\beta)}B(\rh)-2\frac{4-13\beta+7\be2+8\be3}
        {\beta(1-2\beta)^2}\ln(2\beta) \nonumber \\
   & & +\frac{1}{\beta}\Big(\frac{\pi^2}{6}-\dilog(1-2\beta)\Big)+4\frac{1-
     \be2}{\be2}\Big(\frac{\pi^2}{3}+\dilog(1-2\beta)-2\arcsin^2\sqrt{\rh}\Big)
\end{eqnarray}
where $\beta=(1-M_H^2/M^2)$ and $B(\rho)=\sqrt{\rho(1-\rho)}\arcsin\sqrt{\rho
}$. In the same limit one obtains
\begin{eqnarray}
   \delta k^\eta_{ZH} & \to & 2\frac{1-4\beta}{1-2\beta}-\frac{\pi^2}{2}-\frac{
       8(2-\beta)}{(1-\beta)\be2}B(\rh)-4\Big(\frac{2}{\beta}+\frac{1-\beta}
       {(1-2\beta)^2}\Big)\ln(2\beta) \nonumber \\ && +
       2\frac{2-\beta}{\be2}\Big(\frac{\pi^2}{3}+\dilog(1-2\beta)-
       2\arcsin^2\sqrt{\rh}\Big)
\end{eqnarray}

The limiting behavior of $\delta k^\psi_{\gamma H}$ is well known:
\begin{eqnarray}
  \delta k^\psi_{\gamma H} & = & -4-12\ln2-\frac{\pi^2}{4}\qquad(\beta\to1)\\
  \delta k^\psi_{\gamma H} & = & -\frac{8\pi}{3}\frac{1}{\sqrt{\beta}}
           +8-\frac{\pi^2}{2}  \qquad(\beta\to0)
\end{eqnarray}
For $\delta k^\eta_{ZH}$, one obtains in the corresponding limit
\begin{eqnarray}
  \delta k^\eta_{ZH} & = & -2-8\ln2 \qquad(\beta\to1) \\
  \delta k^\eta_{ZH} & = & -\frac{8\pi}{3}\frac{1}{\sqrt{\beta}}
           +6-\frac{\pi^2}{2} \qquad (\beta\to0)
\end{eqnarray}

\subsection{The decays into $W^+W^-$}

Although the decays into $W^+W^-$ seem very similar to the decays into
two $Z$, important differences arise: two distinctively different
isospin assignments must be considered: case 1 for an isosinglet quark and
case 2 for a standard model quark with $|I_3|=1/2$ (isodoublet). In case 2
the propagation of the isospin partner of the quark forming the bound state
enters the diagram. Its mass is generically denoted by $m_T$. For an
isosinglet quark (case 1), only diagram (a) is possible, and we obtain
\begin{eqnarray}
   R^\eta_{WW} & = & 0 \nonumber \\
   R^\psi_{WW} & = & \frac{1}{Q^2}\frac{\be3}{48(1-\rz)^2}
    (1+20\rw+12\rw^2)\left(\frac{M_Z}{M_W}\right)^4
\end{eqnarray}

In case 2, diagrams (a) and (c) contribute, and we find
\begin{eqnarray}
    R^\eta_{WW} & = &
\frac{1}{(Q\sin\Theta)^4}\frac{\be3}{8(1+4\rho_T-4\rw)^2}\\
    R^\psi_{WW} & = & \frac{1}{(Q\sin\Theta)^4}\frac{\be3}{768\rw^2}
       \Bigg[(2I_3)^2\frac{(1+20\rw+12\rw^2)(1-4c\rz)^2}{(1-\rz)^2}\nonumber \\
    & & -16I_3\rw\frac{(1-4c\rz)(5+6\rw)}{(1+4\rho_T-4\rw)(1-\rz)}
        +16\rw^2\frac{2-\rw}{(1+4\rho_T-4\rw)^2}\Bigg]
\end{eqnarray}
where
\[ \beta=\sqrt{1-4\rw}~,~\rho_T=\frac{m_T^2}{M^2}~,~c=2I_3
   \sin^2\Theta_W\]

The dependence on the sign of $I_3$ which is in particular relevant for the
interference term is displayed explicitly\footnote{The sign of this term is in
conflict with the formula given in \cite{top,barger} for d-type quarks.}, so
our result applies to up- as well as to down-type quarks. In the limit of a
large quark mass, $R^\psi_{WW}$ rises $\propto1/\rw^2$, whereas $R^\eta_{WW}$
approaches a constant. This again is a consequence of the equivalence theorem,
which tells us that in the $\psi$ decay both $W$ can be longitudinal. In
contrast both $\eta\to G_WG_W$ and $\eta\to G_WW$ are forbidden. Obviously the
$WW$ mode would dominate $\psi$ decays.

For case 1 (isosinglet) the K-factor is simply constant:
\begin{equation}
   \delta k^\psi_{WW} = 2-\frac{\pi^2}{2}
\end{equation}

\begin{figure}\begin{center}
  \epsfysize 14cm \leavevmode\epsfbox{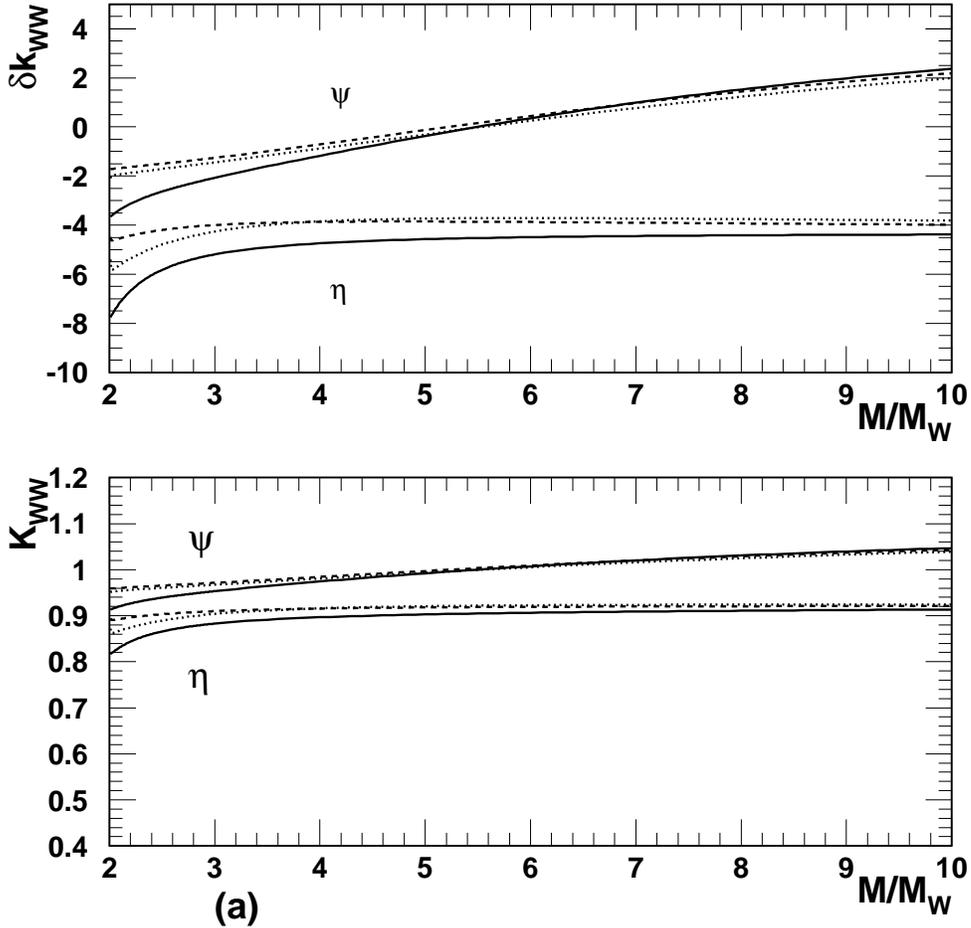}\end{center}
  \caption{$\delta K_{WW}$ and $K_{WW}$ for down-type quarks
            and $m_T-m=0,50,80$GeV (solid, dashed, dotted).\label{kww}}
\end{figure}

The K-factors for case 2 (standard model) are again quite lengthy and are
listed in the appendix. Formally $\delta k^\eta_{WW}=\delta k^\eta_{ZZ}$ if we
set $m_T=m,~v=a$. Fig.~\ref{kww} shows the two functions for three different
values of $m_T$.

\section{Large Quarkonium mass and approximate formul\AE}

The K-factors approach quickly their asymptotic behavior for $\beta$ towards
zero or 1. Since additional quarks are presumably rather heavy, it often will
be sufficient to take the limiting values $\rho_X\to0$. They are given in table
\ref{limits}. In the case of $WW$ a possible mass splitting between the
constituent quark and its isospin partner is neglected.

Up to obvious coupling constants, the ratio between $R^\psi_{ZZ}$ and
$R^\psi_{Z\gamma}$ approaches 1/2 in this limit, a consequence of the
statistical factor for identical particles in the $ZZ$ case. The K-factors
therefore approach the same value. Also $K^\psi_{ZH}$ and $K^\psi_{\gamma H}$
become equal, and the asymptotic value of $R^\eta_{ZZ}$ remains unaffected by
the QCD corrections. The difference between $K^\eta_{ZZ}$ and $K^\eta_{WW}$
only arises from their different couplings to left- and right-handed quarks
and the appearance of $m_T$ in the $WW$ case. It should be mentioned that
$K^\psi_{WW}$ -- for a down-type quark -- approaches its asymptotic value very
slowly, so in this case the full result given in the appendix should be used.

The following approximations are valid for the full kinematical range:
\begin{eqnarray}
  \delta k^\eta_{ZZ} &\approx& -5.84-4.36\rz-47.73\rz^2 \\
  \delta k^\psi_{ZZ} &\approx& \frac{\sqrt{3}\pi}{1-4\rz}-13.66+3.09\rz
         -35.41\rz^2 \\
  \delta k^\eta_{Z\gamma} & \approx& -0.642\rz-0.473\rz^2 \\
  \delta k^\psi_{Z\gamma} & \approx& -\frac{8\pi}{3}\frac{1}{\sqrt{1-\rz}}+0.1
\end{eqnarray}
They reproduce the exact result within an error of less than 1\% even close to
threshold.

\begin{center}\begin{minipage}{11cm}\begin{table}
  \caption{Asymptotic values of the $\delta k$-factors for $\rho_X\to0$.
           Note that in order to obtain the full corrections, these numbers
           have to be multiplied by $\alpha_S(M)\cdot C_F/2\pi$.\label{limits}}
  \begin{tabular}{lcd}
     K-factor & $\rho_X\to0$ & numerical value\\ \hline
     $\delta k^\eta_{ZZ}$ & $-\frac{2a^2}{v^2+a^2}(\pi^2-8\ln 2)$ &
        -5.84\tablenotemark[1] \\
     $\delta k^\eta_{Z\gamma}$ & 0 & 0.0 \\
     $\delta k^\eta_{ZH}$ & $-2-8\ln 2$ & -7.55 \\
     $\delta k^\eta_{WW}$ & $-\pi^2+8\ln 2$ & -4.32 \\
     $\delta k^\psi_{ZZ}$ &  $-\frac{3\pi^2}{4}+2-4\ln 2$ & -8.17 \\
     $\delta k^\psi_{Z\gamma}$ & $-\frac{3\pi^2}{4}+2-4\ln 2$ & -8.17\\
     $\delta k^\psi_{ZH}$ &  $-\frac{\pi^2}{4}-4-12\ln 2$ & -14.79\\
     $\delta k^\psi_{\gamma H}$ & $-\frac{\pi^2}{4}-4-12\ln 2$ & -14.79\\
     $\delta k^\psi_{WW}$ & $2(1+2I_3)-\frac{\pi^2}{2}(1+4I_3)$ &
        4.93\tablenotemark[1]$^,$\tablenotemark[2] \\
       & & -10.80\tablenotemark[3]$^,$\tablenotemark[2]
  \end{tabular}
  \tablenotetext[1]{down-type standard model quark assumed}
  \tablenotetext[2]{equal masses for the isospin partners assumed}
  \tablenotetext[3]{up-type standard model quark assumed}
\end{table}\end{minipage}\end{center}

\section{Summary}

The complete evaluation of the QCD corrections to annihilation decays of
superheavy S-state quarkonia into elektroweak (gauge) bosons has been
presented. Nearly all the corrections are negative, with $\psi\to WW$ (d-type
quark) being the only exception. Some of them are sizeable even in the high
mass regime, where all mass scales can be neglected and the K-factors become
simple constants. In particular the corrections to $\psi\to ZH$ and $\gamma H$
are of the order of 30\% in this limit, whereas the others are all below
18\%. The dominance of the mode $\eta\to ZH$ among the $\eta$-decays for a
certain range of Higgs and quarkonium masses is not changed by strong radiative
corrections.

\appendix
\section*{}
\subsection{Generalties}
Before discussing the full analytic results for $K_{ZH}$ and $K_{WW}$,
the following functions have to be introduced:
\begin{eqnarray*}
  Q_{01} & = & \Big(\frac{m^2-m_T^2}{M_H^2}-1\Big)\ln\frac{m_T}{m}-
               2\sqrt{\Big(\frac{m+m_T}{M_H}\Big)^2-1}\sqrt{1-\Big(\frac{m-m_T}
               {M_H}\Big)^2} \\ &&
               \times\arctan\sqrt{\frac{1-\Big(\frac{m-m_T}{M_H}
               \Big)^2}{\Big(\frac{m+m_T}{M_H}\Big)^2-1}} \\
  Q_{02} & = & Q_{01}\Big(M_H\to M_Z\Big) \\
  Q_{12} & = &\frac{m_T^2-q^2}{q^2}\ln\frac{m_T^2-q^2}{m_T^2}-
              \ln\frac{m_T^2}{m^2}\\
  Q_1 & = & m^2C_0\big(m^2,M_Z^2,q^2;0,m,m_T\big)\qquad,\qquad
  Q_2 = m^2C_0\big(m^2,M_H^2,q^2;0,m,m_T\big) \\
  Q_0 & = & m^4D_0\big(m^2,4m^2,m^2,M_H^2,q^2,M_Z^2;\lambda,m,m,m_T)
          +\frac{m^2}{2(m_T^2-q^2)}\Big(\ln\frac{\lambda^2}{m^2}
            +\frac{\pi^2}{2|\vec v|}-2\Big)
\end{eqnarray*}
$\lambda$ is the gluon mass to regulate the infrared and Coulomb divergences
as mentioned in section II, $\vec v$ denotes the quark velocity and the scalar
three and four point one loop integrals $C_0$ and $D_0$ are defined by
\cite{denner}
\begin{eqnarray*}
  \lefteqn{D_0(p_1^2,(p_1-p_2)^2,p_2^2,(p_2-p_3)^2,p_3^2,(p_3-p_1)^2;
   m_0,m_1,m_2,m_3)} &&\\ &&
  =\int\!\frac{d^4l}{i\pi^2}\frac{1}{[l^2-m_0^2][(l+p_1)^2-m_1^2]
  [(l+p_2)^2-m_2^2][(l+p_3)^2-m_3^2]}
\end{eqnarray*}
\begin{eqnarray*}
  C_0(p_1^2,(p_1-p_2)^2,p_2^2;m_0,m_1,m_2) &=&\int\!\frac{d^4l}{i\pi^2}
     \frac{1}{[l^2-m_0^2][(l+p_1)^2-m_1^2][(l+p_2)^2-m_2^2]}
\end{eqnarray*}
In the case of $K_{ZH},~m_T=m$, which makes the expressions more compact, and
for $K_{WW}$ both $M_H$ and $M_Z$ must be replaced by $M_W$.

When taking the limit $\lambda\to0$ the four point function $D_0$ develops an
infrared and a Coulomb singularity which are exactly of the displayed
separately in the definition of $Q_0$, so the latter is finite. When
calculating $Q_0$, it is not necessary to work out the full expression for
$D_0$, because use can be made of the special kinematical situation which
results in relations between the propagators appearing in the denominator:
\begin{eqnarray*}
  2\lambda^2 & = &[(l+p_q)^2-m^2]+[(l-p_q)^2-m^2]-2[l^2-\lambda^2] \\
  2l^2 & = & [(l+p_q)^2-m^2]+[(l-p_q)^2-m^2]
\end{eqnarray*}
where $p_q$ is the momentum carried by the incoming quark and thus, in our
approximation, $-p_q$ the momentum of the antiquark. These identities
can be used to rewrite $D_0$:
\begin{eqnarray*}
  D_0 & = & \frac{1}{2\lambda^2}\Big( C_0(m^2,M_H^2,(p_q-p_Z)^2;\lambda,m,m_T)
            -C_0(m^2,M_H^2,(p_q-p_Z)^2;0,m,m_T) \\
      &   & +C_0(m^2,M_Z^2,(p_q-p_Z)^2;\lambda,m,m_T)
            -C_0(m^2,M_H^2,(p_q-p_Z)^2;0,m,m_T) \Big)
\end{eqnarray*}
and the calculation is reduced to the problem of expanding three point
functions, with the result:
\begin{eqnarray}
  Q_0(ZH) & = & \frac{1}{2z}\Big[d(\rz)+d(\rh)\Big] \qquad\mbox{with} \\
  d(\rho_X) & = & \Big[\frac{4\rho_X+z-2\beta}{z-2\beta}-\frac{z}{2-z-2\rho_X}
       \Big]\ln\frac{2\beta}{z}-\frac{8}{z}B(\rho_X)+\frac{2(1-z)}{z}\Bigg[
       \frac{1}{4\rho_X+z-2+2\beta} \nonumber \\ &&
       \Big[2\frac{4\rho_X-1+2\beta/z}{4\rho_X+z-2-2\beta+4\beta/z}
       \ln\frac{4\beta(1-1/z)}{z+2\beta+4\rho_X-2}-\ln\frac{4\beta(1-1/z)}
       {z(z+2\beta+4\rho_X-2)} \Big] \nonumber \\ &&
       +\frac{1}{z-2\beta+4\rho_X-2}\Big[\frac{2(1-2\beta/z)}{z-2\beta+4\rho_X
       -2+4\beta/z}\ln\frac{4\beta/z}{2\beta+2-4\rho_X-z} \nonumber \\ &&
       -\frac{z+2-4\rho_X-2\beta}{z-2+4\rho_X+2\beta}
       \ln\frac{4\beta}{2\beta+2-4\rho_X-z}\Big]\Bigg]\qquad z=2(1-\rz-\rh)
       \nonumber \\
  Q_0(WW) & = & -\frac{1}{z}\ln\frac{z}{t}-\frac{\sqrt{4t^2-z^2}}{z^2}\Bigg[
            \arctan\frac{(2-z)(2t^2-z)+\beta(4t^2-z^2)}
            {\sqrt{4t^2-z^2}(2t^2-z-\beta(2-z))}\nonumber \\
      &&    +\arctan\frac{(2t^2-z)^2-\beta(4t^2-z^2)}{\sqrt{4t^2-z^2}(2t^2-z)
            (1+\beta)} \Bigg]
             \qquad z=t^2+\be2,~t\le 1+\frac{M_W}{m}
\end{eqnarray}

To demonstrate the finiteness of $Q_0$, a third relation can be used, which
allows to explicitely split off the divergent part of $D_0(\lambda\to0)$:
\[ l^2+2l(p_q-p_Z)-[(l+p_q-p_Z)^2-m_T^2]=m_T^2-q^2 \]
resulting in
\begin{eqnarray*}
  D_0(\lambda\to0) & = & -\frac{1}{m_T^2-q^2}C_0(m^2,4m^2,m^2;\lambda\to0,m,m)
     \\ &&
     +\frac{1}{m_T^2-q^2}\int\!\frac{d^4l}{i\pi^2}\frac{l^2+2l(p_q-p_Z)}{l^2
     [(l+p_q)^2-m^2][(l-p_q)^2-m^2][(l+p_q-p_X)^2-m_T^2]}\\
    &=& \frac{1}{2m^2(m_T^2-q^2)}\Big(\ln\frac{\lambda^2}{m^2}+\frac{\pi^2}{2
      |\vec v|}-2\Big)+\mbox{finite integral}
\end{eqnarray*}
This equation, however, is not of much practical use because the calculation
of the remaining finite integral by standard methods would re-introduce
infrared divergencies when expanding it into two- and three-point functions.

\subsection{K-factors for the $ZH$ mode}
With these ingredients, the missing K-factors can be calculated. First the
results for $K_{ZH}$:
\begin{eqnarray}
  \delta k^\eta_{ZH}  &=& 2\frac{3-4\rz-4\rh}{1-2\rz-2\rh}-\frac{\pi^2}{2}+
   \left(\frac{\rz-\rh}{\be2}+\frac{1}{1-2\rz-2\rh}\right)2Q_{12}
   +4\frac{1-\rz}{\be2}Q_{01} \nonumber \\ &&
   +4\frac{\rz}{\be2}Q_{02}-4\frac{1-(\rh-\rz)^2}{\be2}Q_1
   -8\frac{\rz(1-\rz+\rh)}{\be2}Q_2+8\frac{(1-\rz-\rh)^2}{\be2}Q_0
\end{eqnarray}

The only form allowed for the amplitude for $\eta\to ZH$ is proportional to
$\epsilon_Z\cdot P_{q\bar q}$. Hence the radiative corrections can trivially
be written as a multiple of the Born result and the K-factor assumes a fairly
simple form. However, the same consideration does not apply to $\psi\to ZH$
where two quite different and more complicated structures are present for the
amplitudes on the tree level. Therefore the corrections cannot be written as a
single `compact' K-factor. However, the explicit calculation shows that every
amplitude corresponding to a given graph of Fig.~\ref{qcds} can be split into
two parts that are each proportional to one of the Born amplitudes. Hence,
after multiplication with the latter and after spin summation, the three
expressions (\ref{zhs}) are recovered and `partial K-factors' can be read off
(with $z=2(1-\rz-\rho_H)$):
\begin{eqnarray}
  \delta k_s & = & \frac{1+\rz-2z\rz}{\rz(1-z)}-\frac{\pi^2}{2}-\frac{(1-\rz)
     (2-3z)}{\rz z(1-z)}Q_{12}+2F_1(\rz,\rho_H) \\
  \delta k_{st} & = & \frac{3(1-z)+\frac{1}{\rz}}{2(1-z)}\!-\!\frac{\pi^2}{2}\!
     -\!\frac{\frac{2-3z}{\rz}+(2-z-z^2)}{2z(1-z)}Q_{12}\!+\!F_1(\rz,\rho_H)
     \!+\!F_2(\rz,\rho_H)\!+\!\frac{\rz F_3(\rz,\rho_H)}{12} \nonumber \\
  && +\frac{z}{6\rz(1+\rz-\rho_H)}\Big[\frac{
     3\rz+\rho_H}{2}Q_{12}+(\frac{z}{4}-1)Q_{01}-\rz Q_{02}+8\rz(z+3\rz)
     Q_0 \nonumber \\
  && -\frac{4(3\rz-1+\rho_H)+z(1+\rz-\rho_H)}{4}Q_1+\frac{8\rz+z(1+\rz-\rho_H)
     -2\be2}{4}Q_2\Big] \\
  \delta k_t & = & \frac{2-z}{1-z}-\frac{\pi^2}{2}-\frac{4-4z-z^2}{z(1-z)}
     Q_{12}+2F_2(\rz,\rho_H)+\frac{\rz^2}{\be2+6\rz}F_3(\rz,\rho_H)\nonumber \\
  && +\frac{z}{\be2+6\rz}\Big[Q_{12}-Q_{01}+2\rz Q_2+(\frac{z}{4}-1)Q_1
     +2(2\be2+12\rz+z)Q_0\Big]
\end{eqnarray}
where
\begin{eqnarray*}
  F_1(\rz,\rho_H) & = & -\frac{1-\rz}{\rz\be2}\Big((1+\rho_H-\rz)
     Q_{01}-2\rz Q_{02}+\frac{5\rz-1-\rho_H}{2}Q_{12} \\
  && +[(\rho_H-\rz)^2-1]Q_1+2\rz(1+\rho_H-\rz)Q_2\Big) \\
  F_2(\rz,\rho_H) & = & 1-\frac{4\rho_H}{\be2}Q_{01}+\frac{3\rz(1+\rho_H
     -\rz)}{\be2}Q_{02}+4\Big[1-\rho_H+\frac{\rho_H(1-\rho_H+\rz)}{\be2}
     \Big]Q_1 \\
  && +\frac{\be2+4\rz^2-4\rz+\frac{z}{8}(4\rz+2-z)+2\rho_H}
     {\be2}Q_{12}-\Big[1+4\rz+\frac{12\rz\rh}{\be2}\Big]Q_2 \\
  F_3(\rz,\rho_H) & = & \frac{z(z/4-\rz)}{\rz^2(1-z)}+\frac{1}{\be2}
     \Big[[12(\rz-1-\rho_H)-4\frac{\be2}{\rz}]Q_{02}+72\frac{1-\rz}{1-z}Q_{12}
     \\ && +\frac{\be2(64\rz-20z\rz+z^2)+z\rz(96\rz^2-216\rz-12+24z\rz)}
     {4\rz^2(1-z)}Q_{12} \\
  && +2[\frac{\be2}{\rz}(6-z)-12z+24(1-\rz)]Q_2\Big]
\end{eqnarray*}

The corrected ratio is then given by
\[ R^\psi_{ZH} = \Big(R^\psi_{ZH}\Big)_{Born}+\frac{\alpha_SC_F}{2\pi}\Big(
     \delta k_sR^\psi_{ZH,s}+\delta k_{st}R^\psi_{ZH,st}+ \delta k_t
     R^\psi_{ZH,t}\Big)
\]

It should be stressed, however, that the notation does not imply that for
example $\delta k_s$ is the correction induced by the s-channel diagrams
fig. \ref{qcds}(a,b) only. In fact, it contains part of the t-channel
contribution and must be considered in the limit $M_Z\to0$ to obtain the
correct result for $K^\psi_{\gamma H}$, although on the Born level only the
t-channel diagrams are possible in this limit.

\subsection{K-factors for the $WW$ mode}
The last missing K-factors are those for the decay modes $\psi\to WW$ and
$\eta\to WW$. They read
\begin{eqnarray}
 \delta k^\eta_{WW} &=&
  \frac{\be2-t^2}{\be2}-\frac{\pi^2}{2}-\frac{\be4(1-\be2-2t^2)-4\be2t^2-
    \be2t^4-t^4}{\be4}\ln\frac{t^2+\be2}{t^2}+\be2\ln(t^2) \nonumber \\ &&
    -(2-t^2-\be2)Q_{02}+2\frac{t^2+\be2}{\be2}Q_2+2\frac{\be4-t^4}{\be2}Q_0
\end{eqnarray}
For $R^\psi_{WW}$ there are again three partial K-factors:
\newcommand{\nn}{7+\beta^2}
\newcommand{\nnn}{2\be2(1-\be2)(13-3\be2)}
\begin{eqnarray}
   \delta k_s &=& 2-\frac{\pi^2}{2} \\
   \delta k_t &=& -\frac{(\be2+t^2)(7+t^2-2\be2)}{\be2(\nn)}-\frac{\pi^2}{2}
      -\frac{t^2(35\be2+4\be4+7t^2+\be2t^2+t^4)}{\be2(\nn)(\be2+t^2)}\ln(t^2)
      \nonumber \\
   &&
+\Big[\frac{\be2(\be2-4t^2)-t^4}{\be2(\be2+t^2)}+\frac{6t\be2-11\be4+2\be6
	+2t\be4+2t^2\be4-t^4}{\be2(\nn)}\Big]Q_{12} \nonumber \\
   && +2\frac{4\be4-4\be2-\be6+3t-2t\be2-t\be4+4t^2-t^2\be2-t^2\be4-t^4}
	{\be2(\nn)}Q_{02} \nonumber \\
   &&
+\frac{4\be6\!+\!t^2(14\!-6t-13t^2+3t^4)\!+\!\be2(14\!+6t-24t^2-2t^3+2t^4)\!
	-\!\be4(3-2t-3t^2)}{\be2(\nn)}Q_{2} \nonumber \\
   && +\frac{14\be2-9\be4+4\be6-t^2(14+8\be2-7\be4-5t^2-2\be2t^2+t^4)}
      {\be2(\nn)}(\be2+t^2)Q_0 \\
   \delta k_{st} &=& -\frac{\pi^2}{2}-\frac{(\be2+t^2)(13-13\be2+3t^2+2\be4
	-t^2\be2)}{\nnn}\nonumber \\
   && -\frac{t^2(130\be2-142\be4+29\be6+26t^2-8t^2\be2+8t^2\be4+6t^4+3t^4\be2)}
	{\nnn(\be2+t^2)}\ln(t) \nonumber \\
   && +\Big[\frac{\be2(13+14t-2t^2+5t^4)-\be4(27+16t-11t^2)+2\be6(8-t)-13t^2
	-3t^4}{\nnn} \nonumber \\
   && -\frac{2t^2}{t^2+\be2} \Big]Q_{12}
	+\frac{7t+t^4-10\be2-8t\be2+4t^2\be2+9\be4+t\be4}{\be2(13-3\be2)}Q_{02}
	\nonumber \\
   && +\Big[\frac{\be2(26+14t-82t^2+16t^3-17t^4-3t^6)-\be4(58+16t-37t^2+2t^3
	+t^4)}{\nnn} \nonumber \\
   && +\frac{\be6(27+2t+t^2)-\be8+26t^2-14t^3+16t^4-3t^6}{\nnn}
+\frac{\be4-t^4}
	{2\be2}\Big]Q_2 \nonumber \\
   && +\frac{\be2(26+10t^2+5t^4+t^6)-\be4(51-17t^2)+\be6(25-3t^2)-2\be8-26t^2
      -3t^4+t^6}{2\be2(1-\be2)(13-3\be2)} \nonumber \\
   && \times(\be2+t^2)Q_0
\end{eqnarray}

The corrected rate can be obtained analogously to $R^\psi_{ZH}$.

\end{document}